\documentclass[conference]{IEEEtran}
\IEEEoverridecommandlockouts
\usepackage{cite}
\usepackage{amsmath,amssymb,amsfonts}
\usepackage{algorithmic}
\usepackage{graphicx}
\usepackage{textcomp}
\usepackage{xcolor}
\usepackage{hyperref}
\usepackage{float}
\usepackage{url}
\def\BibTeX{{\rm B\kern-.05em{\sc i\kern-.025em b}\kern-.08em
    T\kern-.1667em\lower.7ex\hbox{E}\kern-.125emX}}

\usepackage{fancyhdr}
\begin{document}

\title{PARK: Parkinson's Analysis with Remote Kinetic-tasks
\thanks{This project was funded by \emph{Gordon \& Betty Moore Foundation} and \emph{Google}.}
}


\author{Md Saiful Islam$^*$,
        Sangwu Lee$^*$,
        Abdelrahman Abdelkader,
        Sooyong Park,
        and Ehsan Hoque\\
        Department of Computer Science, University of Rochester, Rochester, United States\\
        mislam6@ur.rochester.edu, \{slee232, aabdelka, spark180\}@u.rochester.edu, and mehoque@cs.rochester.edu
\thanks{$^*$Equal contribution.}
}

\maketitle
\thispagestyle{fancy}

\begin{figure*}
\centering
\includegraphics[width=0.75\textwidth]{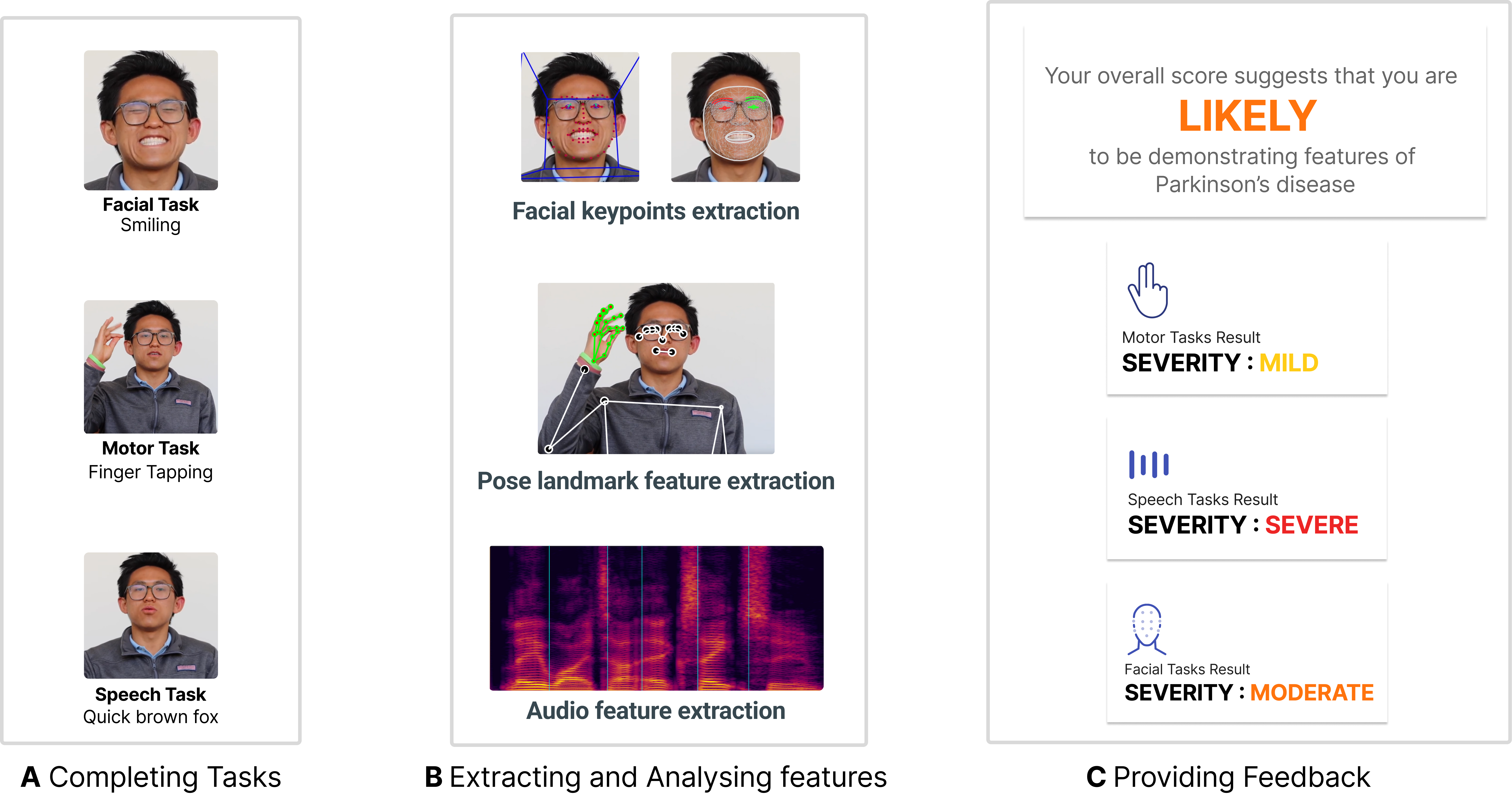}
\caption{\label{fig:park_pipeline}An end-to-end framework for Parkinson's disease screening.}
\end{figure*}

\begin{abstract}
We present a web-based framework to screen for Parkinson's disease (PD) by allowing users to perform neurological tests in their homes. Our web framework guides the users to complete three tasks involving speech, facial expression, and finger movements. The task videos are analyzed to classify whether the users show signs of PD. We present the results in an easy-to-understand manner, along with personalized resources to further access to treatment and care. Our framework is accessible by any major web browser, improving global access to neurological care.
\end{abstract}

\begin{IEEEkeywords}
Parkinson's disease, screening, user-centric framework
\end{IEEEkeywords}

\section{Introduction}
Diagnosis of Parkinson's disease (PD), the fastest-growing neurological disorder~\cite{dorsey2018emerging}, is challenging. Currently, there is no reliable biomarker for detecting PD~\cite{yang2022artificial}, and the only way to get diagnosed is to see a neurologist and go through clinical procedures that may span multiple visits. Imagine someone in their sixties living in a remote area with impaired driving and cognitive ability caused by PD. This person could be reluctant to seek clinical diagnosis until symptoms have worsened significantly. The scenario may be even worse for someone living in a developing or underdeveloped country. In India, there were only 1200 neurologists for more than 1.3 billion people in  2013~\cite{khadilkar2013neurology}. On average, there is only one neurologist for more than 3 million African population, and 21 countries have less than five neurologists~\cite{kissani2022does}. Under the current circumstances, a significant part of the world population could be living with Parkinson's disease without realizing it. When diagnosed late, the disease could have progressed significantly, limiting the usefulness of medications available to manage involuntary tremors caused by PD.

Telemedicine platforms for Parkinson's disease have gained substantial attention from researchers. For example, Machine Medicine Technologies developed a platform named KELVIN~\cite{rupprechter2021clinically} that allows neurologists to record their patients completing various tasks and assign relevant MDS-UPDRS scores. Later, KELVIN uses these scores as the ground truth and builds machine learning models to analyze similar videos automatically. However, the platform is built for clinical providers only and is inaccessible to people without a provider. The recent advancement in affective computing has made it possible to automatically analyze recorded audio and videos to diagnose PD from multiple aspects (\cite{PARK,ALI-headpose,videoPARK}). Prior research attempted automated speech analysis, where individuals utter ``aa...h'' and sustain their tone steadily as long as possible (\cite{wroge2018parkinson, dubey2015echowear}). Also, researchers developed machine learning models to assess PD by looking at motor symptoms (\cite{ali2020spatio, islam2023using}) and analyzing facial expressions from recorded videos~\cite{jin2020diagnosing}. While these works have advanced the state of computing, many have little utility to current healthcare as they do not generalize well when tested with real-life home-recorded data. Moreover, integrating these technologies with existing clinical care requires careful consideration of several factors, including but not limited to communicating the outcome to patients, an extensive assessment of risk versus benefit, and multiple clinical trials. As a result, automated screening/diagnosis remains inaccessible to the public despite significant progress enabled by AI.

This work takes a step toward a near-term future where advances in affective computing and improved access to computing could screen individuals for Parkinson's disease and guide them to a faster referral to see neurologists and help them monitor the progression of the disease. Anyone with access to an Internet connection can go to our website, perform some tasks involving speech, facial expressions, and motor coordination, and get a preliminary screening assessment as to whether they are showing signs of Parkinsonism. We further guide the user toward seeking appropriate care by providing resources like searching for nearby neurologists, support groups, and care facilities. Figure \ref{fig:park_pipeline} outlines our framework, and a video of the demonstration of this framework is available at \href{https://youtu.be/gXZq5CZLNHA}{\textcolor{blue}{\url{https://tinyurl.com/parktest-acii}}}.

\begin{table*}[!htb]
\centering
\resizebox{1.95\columnwidth}{!}{%
\begin{tabular}{|l|l|l|l|l|}
\hline
\textbf{Task}     & \textbf{Publication year} & \textbf{Treatment population}                                                                             & \textbf{Control population}                                                                       & \textbf{Results}                                                                                                                                                                                                                   \\ \hline
speech            & 2021~\cite{rahman2021detecting}                      & \begin{tabular}[c]{@{}l@{}}262 participants with PD\\ (age: 66.0 (9.2) years;\\ 39\% female)\end{tabular} & \begin{tabular}[c]{@{}l@{}}464 participants\\ (age: 58.0 (14.2) years;\\ 65\% women)\end{tabular} & \begin{tabular}[c]{@{}l@{}}AUC score of 0.753 for distinguishing\\ between individuals with and without\\ PD\end{tabular}                                                                                                          \\ \hline
facial expression & Ongoing                   & 256 participants with PD (44\% female)                                                                    & 803 participants (58\% female)                                                                    & \begin{tabular}[c]{@{}l@{}}AUC score of 0.853 for classifying individuals\\ with self-reported PD from those without PD\end{tabular}                                                                                               \\ \hline
motor             & 2023~\cite{islam2023using}                   & 172 participants with PD (37\% female)                                                                    & 78 participants (36\% female)                                                                     & \begin{tabular}[c]{@{}l@{}}Mean absolute error of 0.58 and Pearson's correlation\\ coefficient of 0.66 compared to the clinician rated \\ severity rating for the finger-tapping task used to \\ assess bradykinesia.\end{tabular} \\ \hline
\end{tabular}%
}

\caption{Performance of the developed models.}
\label{tab:performance}
\vspace{-6mm}
\end{table*}

\vspace{-3mm}

\section{Methods}
The proposed Parkinson's disease (PD) screening system can be disintegrated into three key components:

\begin{itemize}
    \item[] \textbf{Task completion.} In this framework, we guide the participants to complete some standard tasks used to assess PD by neurologists. The tasks involve multiple modalities: speech, facial expressions, and motor coordination. For the speech task, the participant is instructed to utter a pan-gram, ``the quick brown fox jumps over the lazy dog,'' and the speech is recorded using the microphone of their computer. The facial expression involves three tasks where the participant is instructed to mimic three different facial expressions: disgust, smile, and surprise. Finally, for the motor task, the participant needs to tap their thumb finger with their index finger as fast and as big as possible. The tapping is done for both left and right hands separately.
    \item[] \textbf{PD risk assessment model.} Based on data that we have collected, we developed three separate models to assess signs of Parkinson's disease from the speech task, the facial expression tasks, and the motor task. For the speech task~\cite{rahman2021detecting}, we use an XGBoost model trained on standard acoustic features extracted from the recordings for determining the presence of self-reported PD. For the facial expression tasks, we use an ensemble of support vector machines trained on facial landmarks and action units extracted using MediaPipe and OpenFace, respectively. Finally, for the motor task~\cite{islam2023using}, we employ a LightGBM regressor trained on custom hand-movement features to assess the severity of bradykinesia.
    
    \item[] \textbf{Communicating the assessment results to the subject.} We consulted several clinicians, including expert neurologists and PD care providers, to find appropriate language to communicate the predictions of the machine learning models to the participants. Currently, the framework presents an overall likelihood of someone showing signs of Parkinsonism, followed by the severity of symptoms across speech, facial expressions, and motor modalities. Then, based on the assessments, it provides further resources that include searching for nearby neurologists and support groups, diet and exercise suggestions, and links to external resources.
\end{itemize}

\section{Results}
To this end, we evaluated the models that learn from each of the speech, facial expressions, and motor modalities independently. We present a summary of the performance of these models in Table \ref{tab:performance}.

In the near future, we plan to develop a model that aggregates the modality-dependent models and provides a final prediction as to whether someone may have Parkinson's disease.

\section*{Ethical Impact Statement}

An inaccurate PD screening model could potentially be harmful. For example, incorrectly identifying someone to have PD may cause unnecessary stress and financial burden, while missing out on someone having PD may give them a false sense of security. The proposed framework and the demonstration are for a technical audience only and are not intended for clinical use in their current state.

\section*{Acknowledgment}
We acknowledge Rafayet Ali, Raina Langevin, Taylan Sen, Victor Anthony,  Wasifur Rahman, Abdullah Al-Mamun, Tariq Adnan, Raye Liu, and Harshil Ratnu for contributing towards the framework. We also acknowledge our clinical collaboraotrs -- Ray Dorsey, Jamie Adams, and Ruth Schneider.

\end{document}